\documentclass[prd,aps,twocolumn,nofootinbib,preprintnumbers,superscriptaddress]{revtex4-2}

\usepackage{amsmath}
\usepackage{color}
\usepackage{graphicx}
\usepackage{hyperref}
\definecolor{darkblue}{rgb}{0.0,0.0,0.6}
\hypersetup{colorlinks, citecolor=darkblue, linkcolor=darkblue, urlcolor=darkblue}

\usepackage[american]{babel}
\usepackage{microtype}

\renewcommand{\v}[1]{\mathbf{#1}}
\newcommand{\be}{\begin{equation}}
\newcommand{\ee}{\end{equation}}
\newcommand{\Fig}[1]{Fig.~\ref{fig:#1}}
\newcommand{\Eq}[1]{Eq.~(\ref{eq:#1})}
\newcommand{\Tab}[1]{Table~\ref{tab:#1}}
\newcommand{\rhodm}{\rho_{_\text{DM}}}
\newcommand{\g}{\gamma}
\newcommand{\grad}{\nabla}
\newcommand{\order}[1]{\mathcal{O}{(#1)}}
\newcommand{\fm}{\text{fm}}

\newcommand{\mK}{\text{mK}}
\newcommand{\PT}{$\mathcal{P}, \mathcal{T}$}
\newcommand{\eV}{\text{e\kern-0.15ex V}}
\newcommand{\keV}{\mathrm{k}\eV}
\newcommand{\GeV}{\mathrm{G}\eV}
\newcommand{\cm}{\mathrm{cm}}
\newcommand{\GHz}{\mathrm{GHz}}

\begin{document}

\preprint{FERMILAB-PUB-22-654-SQMS-T}
\preprint{SLAC-PUB-17702}

\title{Discovering QCD-Coupled Axion Dark Matter with Polarization Haloscopes}

\author{Asher Berlin}
\email{aberlin@fnal.gov}
\affiliation{Theory Division, Fermi National Accelerator Laboratory, Batavia, IL 60510, USA}
\affiliation{Superconducting Quantum Materials and Systems Center (SQMS), Fermi National Accelerator Laboratory, Batavia, IL 60510, USA}
\author{Kevin Zhou}
\email{knzhou@stanford.edu}
\affiliation{SLAC National Accelerator Laboratory, 2575 Sand Hill Road, Menlo Park, CA 94025, USA}

\begin{abstract}
In the presence of QCD axion dark matter, atoms acquire time-dependent electric dipole moments. This effect gives rise to an oscillating current in a nuclear spin-polarized dielectric, which can resonantly excite an electromagnetic mode of a microwave cavity. We show that with existing technology such a ``polarization haloscope" can explore orders of magnitude of new parameter space for QCD-coupled axions. If any cavity haloscope detects a signal from the axion-photon coupling, an upgraded polarization haloscope has the unique ability to test whether it arises from the QCD axion.
\end{abstract}

\maketitle

\section{Introduction}

The QCD axion is a long-standing, well-motivated dark matter candidate~\cite{Peccei:1977hh,Peccei:1977ur,Weinberg:1977ma,Wilczek:1977pj,Preskill:1982cy,Abbott:1982af,Dine:1982ah} that can also explain why the neutron's electric dipole moment (EDM) is at least $10^{10}$ times smaller than generically expected~\cite{Abel:2020pzs}. It is a pseudoscalar field $a$ defined by its coupling to gluons
\be
\label{eq:gluon_coupling}
\mathcal{L} \supset \theta_a \, \frac{\alpha_s}{8 \pi} \, G^{\mu\nu} \tilde{G}_{\mu\nu}
~,
\ee
where $\theta_a \equiv a / f_a$ and $f_a$ is the axion decay constant. At temperatures below the QCD phase transition, this coupling generates a potential and mass for the axion~\cite{GrillidiCortona:2015jxo}
\be
\label{eq:QCDmass}
m_a = 5.7 \ \mu \eV \times ( 10^{12}  \ \GeV / f_a )
~.
\ee
Over cosmological time, the axion field relaxes towards the minimum of its potential at the parity ($\mathcal{P}$) and time-reversal ($\mathcal{T}$) conserving point $\theta_a = 0$ where the neutron EDM vanishes. Assuming a standard cosmological history and an $\order{1}$ initial misalignment angle, the residual energy in the axion field accounts for the present density of cold dark matter for $m_a \sim (0.5 - 50) \, \mu \eV$~\cite{Borsanyi:2016ksw}. In this case, the local axion field has macroscopic mode occupancy and can thus be described by a classical expectation value,
\be
\label{eq:thetaDM}
\theta_a \simeq \frac{\sqrt{2 \rhodm}}{m_a \, f_a} \, \cos{m_a t} \, \simeq \, 4.3 \times 10^{-19} \, \cos{m_a t}
~,
\ee
oscillating with frequency $m_a / 2 \pi \sim (0.1 - 10) \ \GHz$, where $\rhodm \simeq 0.4 \ \GeV / \cm^3$ is the local dark matter density. 

The direct signatures of QCD axion dark matter are nuclear effects, such as the oscillating neutron EDM~\cite{Pospelov:1999ha},
\begin{align}
\label{eq:dn}
d_n &\simeq \big( 2.4 \times 10^{-3} \ e \ \fm \big) \, \theta_a
~.
\end{align}
Detecting such a small signal is very difficult, but has been addressed by several recent proposals. In some cases, static EDM experiments may be repurposed to constrain slowly oscillating EDMs~\cite{Abel:2017rtm,Roussy:2020ily,Schulthess:2022pbp}. Other potential detection avenues involve nuclear magnetic resonance~\cite{Budker:2013hfa,JacksonKimball:2017elr,Aybas:2021nvn}, spin precession in storage rings~\cite{Chang:2017ruk,Pretz:2019ham,Stephenson:2020jzx,Kim:2021pld,Alexander:2022rmq,JEDI:2022hxa}, atomic and molecular spectroscopy~\cite{Graham:2011qk,Kim:2022ype}, and mechanical oscillations in piezoelectric materials~\cite{Arvanitaki:2021wjk}. However, none of these probes are sensitive at the $\GHz$ frequencies motivated by standard misalignment production of axion dark matter. 

Currently, the most stringent laboratory constraints on axion dark matter at $\GHz$ frequencies come from cavity haloscopes~\cite{Sikivie:1983ip,Wilczek:1987mv}, which rely on the axion's coupling to photons, $\mathcal{L} \supset g_{a \g \g} a F^{\mu\nu} \tilde{F}_{\mu\nu}/4$. In these experiments, axion dark matter produces an effective current $\v{J}_{a \g \g} = g_{a \g \g} \v{B} \partial_t a$ inside a microwave cavity with background magnetic field $\v{B}$, which can resonantly excite a mode of angular frequency $m_a$. While there are many other recent proposals to search for the axion (see Refs.~\cite{Irastorza:2018dyq,Sikivie:2020zpn,Semertzidis:2021rxs} for reviews), the cavity haloscope concept is currently the most well-developed, with many collaborations reporting new results~\cite{ADMX:2018gho,ADMX:2019uok,ADMX:2021nhd,Brubaker:2016ktl,HAYSTAC:2018rwy,HAYSTAC:2020kwv,Lee:2020cfj,Jeong:2020cwz,CAPP:2020utb,Lee:2022mnc,Kim:2022hmg,ADMX:2018ogs,Grenet:2021vbb,TASEH:2022hfm,TASEH:2022vvu,CAST:2020rlf,Alesini:2020vny,McAllister:2017lkb,Quiskamp:2022pks} and some operating near or beyond the standard quantum limit~\cite{ADMX:2021nhd,HAYSTAC:2018rwy,HAYSTAC:2020kwv,Kim:2022hmg}. These experiments are well-motivated, as the axion-gluon coupling of Eq.~\eqref{eq:gluon_coupling} is known to induce an axion-photon coupling. However, their relation is indirect: the coefficient $g_{a \g \g}$ can vary by orders of magnitude within simple models~\cite{Kaplan:1985dv,Cheng:1995fd,DiLuzio:2016sbl,DiLuzio:2017pfr}, and an axion with an electromagnetic coupling is not necessarily the QCD axion. Definitively discovering or excluding the QCD axion thus requires confronting the axion-gluon coupling directly. 

\begin{figure}
\includegraphics[width=\columnwidth]{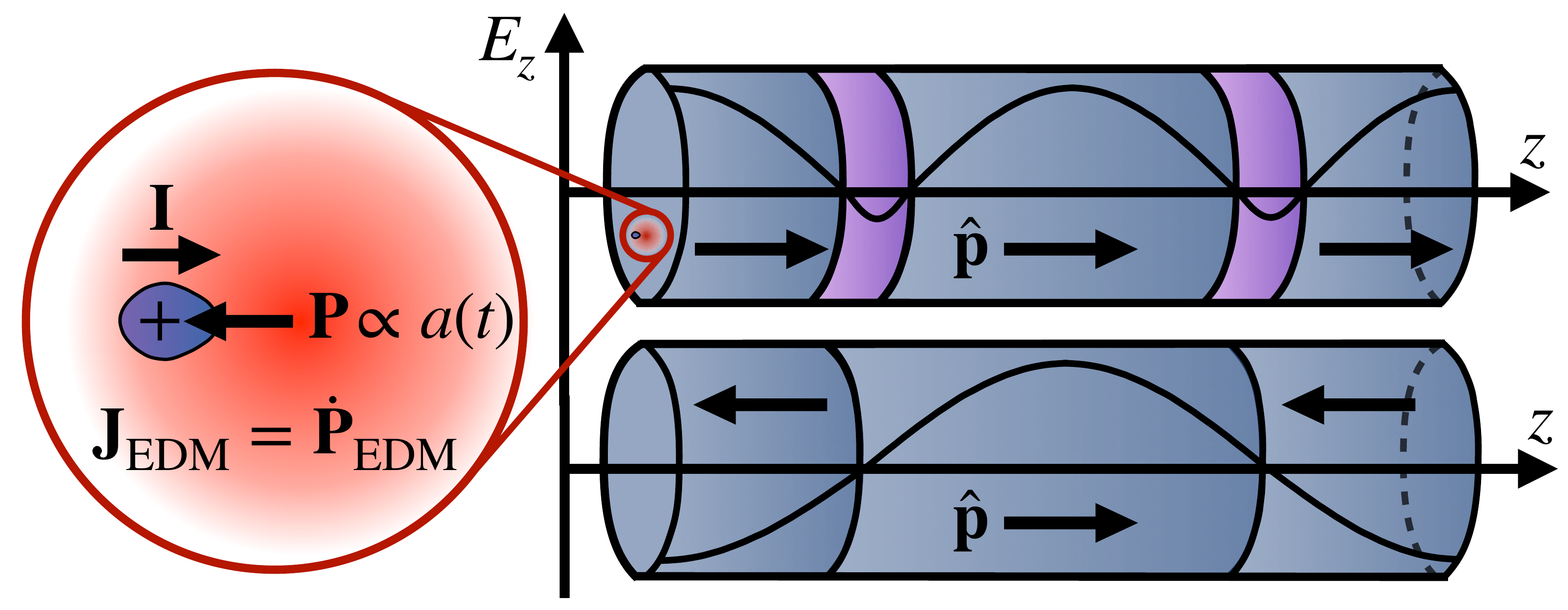}
\caption{Atoms carry EDMs proportional to the axion field (left), aligned with the nuclear spin $\v{I}$. The axion's time variation thus produces a current $\v{J}_\text{EDM}$ in a nuclear spin-polarized dielectric, whose effect can be amplified in a resonant cavity. For higher axion masses, the geometric overlap factor in Eq.~\eqref{eq:eta} can be maximized using layers of inert dielectric (top) or alternating spin polarization (bottom).}
\label{fig:coupling}
\end{figure}

In this work, we present the first method to probe the axion-gluon coupling at GHz frequencies. In the presence of axion dark matter, atoms have oscillating EDMs of magnitude $d_A$ directed along their nuclear spin~\cite{Flambaum:2019emh}, analogous to the neutron EDM in \Eq{dn}. A dielectric thus carries a polarization density $P_\text{EDM} \sim n_A \, d_A$, where $n_A$ is the density of nuclear spin-polarized atoms. A time-varying polarization induces a physical electromagnetic current $\v{J}_\text{EDM} = \partial_t \v{P}_\text{EDM}$, which can be resonantly amplified by placing the dielectric in a microwave cavity with a mode of angular frequency $m_a$. We call this system, depicted in Fig.~\ref{fig:coupling}, a polarization haloscope.

To quickly estimate its potential, we may compare the current in a polarization haloscope to that produced in a typical cavity haloscope. For the benchmark DFSZ model, where $g_{a \g \g} \simeq 0.87 \times 10^{-3} / f_a$~\cite{Zhitnitsky:1980tq,Dine:1981rt}, the ratio is
\be
\frac{J_\text{EDM}}{J_{a \g \g}} \simeq 10^{-3} \times \frac{d_A}{d_n} \, \bigg( \frac{n_A}{5 \times 10^{22} \ \cm^{-3}} \bigg) \, \bigg(\frac{8 \ \text{T}}{B} \bigg)
~,
\ee
which suggests that the signal in a cavity haloscope is larger. Furthermore, $J_{\text{EDM}}$ is more difficult to calculate, as it depends sensitively on nuclear, atomic, and material properties. For these reasons, the polarization haloscope idea was briefly raised and discarded thirty years ago~\cite{PhysRevD.42.1847}. However, the rapid recent progress in cavity haloscopes motivates a thorough analysis of its potential. In section~\ref{sec:EDMs} we show that $d_A \sim d_n$ can be achieved for certain atoms. We then consider the factors necessary to develop an effective polarization haloscope, such as cavity design (section~\ref{sec:cavity}), material choice (section~\ref{sec:material}), and nuclear spin polarization (section~\ref{sec:spin_pol}). We estimate experimental sensitivity in section~\ref{sec:sensitivity} and conclude in section~\ref{sec:conclusion}, laying out a path towards reaching the QCD axion.  


%

\section{Axion-induced EDMs}
\label{sec:EDMs}

The dominant nuclear contribution to the EDM of an atom with atomic number $Z$ arises from the \PT-violating piece of the effective nuclear electric potential~\cite{Khriplovich:1997ga,Flambaum:1984fb,Engel:1999np,Auerbach:1996zd,Spevak:1996tu,Ginges:2003qt}
\be
\label{eq:PhiEff}
\phi_N^{(\text{eff})} (\v{x})  = \Big( 1  + \frac{1}{Z e} \, \v{d}_N \cdot  \grad  \Big)\, \phi_N (\v{x})
~,
\ee
which includes the usual electric potential $\phi_N$ of the nucleus and the response of the atomic electrons to the nuclear EDM $\v{d}_N$. The leading \PT-violating term in a multipole expansion of $\phi_N^{(\text{eff})}$ is the dipole, but it simply vanishes, in accordance with Schiff's theorem~\cite{Schiff:1963zz} which states that the nuclear EDM is efficiently screened by the atomic electrons. The next \PT-violating term is the octupole. Its traceless part corresponds to an electric octupole moment, whose effects are suppressed by the centrifugal barrier near the nucleus~\cite{Khriplovich:1997ga}. The traceful part yields the dominant contribution to the atomic EDM and is described by the Schiff moment~\cite{Flambaum:1984fb}, 
\be
\label{eq:Schiff1}
\v{S} = \frac{1}{10} \, \int d^3 \v{x} ~ \rho_N(\v{x}) \, r^2 \,  \Big(\v{x} - \frac{5}{3} \, \frac{\v{d}_N}{Z e}  \Big)
~,
\ee
where $\rho_N$ is the nuclear charge density; $\v{S}$ sources a \PT-violating electric field that polarizes the atomic electrons, perturbing the electronic Hamiltonian by
\be
\label{eq:HSchiff}
V_S = -  \sum_{i=1}^Z \, e \, \v{S} \cdot \grad \delta^3(\v{x}_i)
~,
\ee
where the nucleus is at the origin. The interaction $V_S$ mixes opposite parity states, which to first order in perturbation theory gives rise to a non-vanishing atomic EDM, parallel to the nuclear spin $\v{I}$, of the form
\be
\label{eq:StoEDM0}
\v{d}_A \simeq \sum_n ~ \frac{\langle n | V_S | 0\rangle \, \langle 0 |  \v{D}  | n \rangle}{E_n - E_0} + \text{h.c.}
~,
\ee
where $|n \rangle$ are atomic states of energy $E_n$ and $\v{D} = - \sum_{i=1}^Z e \, \v{x}_i$ is the atomic EDM operator. The result scales as $d_A \propto Z^2 S$, with a moderate relativistic enhancement for the heaviest nuclei. Scaling numeric results for ${}^{225}\text{Ra}$ from Refs.~\cite{Dzuba:2002kg,Dzuba:2009kn,Dzuba:2007zz,Flambaum:2019tym} yields 
\be
\label{eq:StoEDM}
d_A \simeq - \big(0.27 \times 10^{-3} \ e \ \fm \big) ~ \langle S_z \rangle / (e \ \fm^3)
\ee
for ${}^{161}\text{Dy}$, with values within $20 \%$ for the other nuclei we will consider below. Here, $\langle S_z \rangle$ is the lab-frame expectation value of the Schiff moment directed along the nuclear spin for a maximally-polarized nucleus, $M = I$~\cite{Spevak:1996tu}. 

In perturbation theory, the Schiff moment is
\be
\label{eq:Schiff0}
\langle S_z \rangle \simeq \sum_n ~ \frac{\langle n | V_{\mathcal{P} \mathcal{T}} | 0\rangle \, \langle 0 |  S_z  | n \rangle}{E_n - E_0} + \text{h.c.}
~,
\ee
where $| n \rangle$ are nuclear states of energy $E_n$ and $V_{\mathcal{P} \mathcal{T}} \propto \theta_a$ is the axion's \PT-violating modification to the pion-mediated internucleon interaction. For a typical spherical nucleus with mass number $A$ and radius $R_0 \simeq (1.2 \ \fm) \, A^{1/3}$, we expect~\cite{Flambaum:1984fb,Khriplovich:1997ga}
\begin{align}
\langle n | V_{\mathcal{P} \mathcal{T}} | 0\rangle &\sim (10^{-2} \, \theta_a/m_n \, R_0) ~ (A / m_\pi^2 \, R_0^3)
~, \\
\langle 0 |  S_z  | n \rangle &\sim e R_0^3
~, \\
E_n - E_0 &\sim A / m_\pi^2 R_0^3
~,
\end{align}
which yields the parametric estimate
\be
\label{eq:Schiffspherical}
\langle S_z \rangle \sim 10^{-2} \, \frac{e R_0^2}{m_n} \, \theta_a \sim (0.1 \times e \ \fm^3) \, \theta_a  \bigg( \frac{A}{10^2} \bigg)^{\frac{2}{3}}~,
\ee
in agreement with detailed calculations~\cite{Flambaum:1984fb,Khriplovich:1997ga,Dmitriev:1994mc,deVries:2020iea,Flambaum:2014jta,Haxton:1983dq,Griffiths:1991mk,Arvanitaki:2021wjk}.

This yields only a small atomic EDM, $d_A \ll d_n$, but for nonspherical nuclei there can be a large intrinsic Schiff moment $S_\text{int}$ in the body-fixed frame. Evaluating \Eq{Schiff1} gives $S_{\text{int}} \propto \beta_2 \beta_3 \, Z e R_0^3$, where $\beta_2$ and $\beta_3$ parametrize the quadrupole and octupole deformation of the nuclear radius. The lab-frame Schiff moment is then determined by averaging over nuclear orientations, $\langle S_z \rangle = S_{\text{int}} \langle \hat{n}_z \rangle$ where $\hat{\v{n}}$ is the nuclear axis. A nonzero $\langle \hat{n}_z \rangle$ requires $\mathcal{P}$-violation and is thus proportional to $\theta_a$. It can be calculated perturbatively with an expression analogous to \Eq{Schiff0}, the main difference being that octupole deformations imply states with small energy gaps, $E_n - E_0 \sim 50 \ \keV$. For significantly octupole-deformed nuclei, $\beta_2\sim \beta_3 \sim \order{0.1}$, various numeric factors cancel, leaving~\cite{Dmitriev:1994mc,deVries:2020iea,Flambaum:2014jta,Haxton:1983dq,Griffiths:1991mk}
\begin{align}
\label{eq:Schiff2}
\langle S_z \rangle &\sim 10^{-2} \, \frac{Z e R_0^2}{m_n} \, \theta_a~,
\end{align}
which is crucially enhanced by $Z$ relative to \Eq{Schiffspherical}. Applying \Eq{StoEDM}, we find that for these nuclei,
\be
\label{eq:dArough}
|d_A| \sim \big( \text{few} \times 10^{-3} \big) \ e \ \fm \times \theta_a \,  \bigg( \frac{Z}{10^2} \bigg)^3 \bigg( \frac{A}{10^2} \bigg)^{\frac{2}{3}}
~,
\ee
which, as anticipated above, is comparable to $d_n$.

\begin{table}
\centering
\begin{tabular}{c|ccc}
& ${}^{161}\text{Dy}$ & ${}^{153}\text{Eu}$ & ${}^{155}\text{Gd}$ \\ \hline
estimated $\langle S_z \rangle$ ($e \ \fm^3 \ \theta_a$)~\cite{Dalton:2023kfz} & $4.3$ & $1.0$ & $1.2$ \\ 
estimated $|d_A|$ ($10^{-3} ~ e \ \fm \ \theta_a$) & $1.2$ & $0.25$ & $0.3$ \\ 
natural abundance~\cite{harris2001nmr} & 19\% & 52\% & 15\% \\
metal price ($\$ / \mathrm{ton}$)~\cite{prices} & 300\,k & 30\,k & 30\,k \\
$T \, d f_p/dB |_{B=0} \, (\mK/\text{T})$~\cite{harris2001nmr} & 0.08 & 0.26 & 0.05
\end{tabular}
\caption{Stable nuclei with large axion-induced Schiff moments $\langle S_z \rangle$ and atomic EDMs $d_A$, and their natural abundance and price. We use the last row (equal to $|\gamma| \, (I+1)/3$ where $\gamma$ is the gyromagnetic ratio~\cite{stupic2011hyperpolarized}) to determine the fractional nuclear spin polarization $f_p$ at a temperature $T$ in a magnetic field $B$.}
\label{tab:nuclei}
\end{table}

Most octupole-deformed nuclei are short-lived and thus infeasible to gather in the macroscopic quantities required. Of the nuclei highlighted in Refs.~\cite{Flambaum:2019tym, Dalton:2023kfz, Flambaum:2019kbn}, we identify $^{161}\text{Dy}$, $^{153}\text{Eu}$, and $^{155}\text{Gd}$ as the most promising. They are absolutely stable and, as indicated in \Tab{nuclei}, are inexpensive and expected to possess fairly large axion-induced Schiff moments and atomic EDMs. However, the existence of octupole deformation in these nuclei is not completely settled~\cite{Behr:2022hym}. This work motivates further experimental study. Even if none of these nuclei are octupole deformed, it may still be possible to achieve comparable EDMs via magnetic quadrupole moments, which are enhanced by well-established nuclear quadrupole deformations~\cite{Dalton:2023kfz}. 


\section{Cavity Excitation}
\label{sec:cavity}

The axion field oscillates with a phase offset and amplitude varying over the coherence time $\tau_a \sim Q_a/m_a$, where $Q_a \sim 10^6$. For all axion masses we consider, spatial gradients of the axion field are negligible. The cavity response is therefore very similar to that of a conventional haloscope, with $\v{J}_{a \g \g}$ replaced by $\v{J}_\text{EDM} \simeq m_a \, n_A \, \v{d}_A$. In our case, there is also an associated physical charge density $\rho_\text{EDM} = - \nabla \cdot \v{P}_\text{EDM}$ in the cavity, which produces small electric fields, but it is not of interest because it cannot excite resonant modes~\cite{condon1941forced,smythe1988static,collin1990field}. 

We suppose a portion $V_p$ of the volume $V$ of the cavity is filled with dielectric of fractional nuclear spin polarization $f_p$ along the $\hat{\v{p}}$ direction, so that $n_A = f_p \, n_0$ where $n_0$ is the number density of relevant nuclei. Adapting a standard result~\cite{Kim:2019asb}, the power deposited to the $i^{\text{th}}$ mode of the cavity on resonance, $m_a \simeq \omega_i$, is
\be
\label{eq:signal_power}
P_\text{sig} \simeq m_a  \, (f_p \, n_0 \, d_A)^2  \ (V / \bar{\epsilon}) \, \eta_i^2 \, \min (Q_a , Q_i)
~,
\ee
where $d_A$ is now the time-independent amplitude of the atomic EDM, $Q_i$ is the quality factor of the mode, and the last factor accounts for the spectral width of the axion. The typical dielectric permittivity inside the cavity is $\bar{\epsilon}$, and the geometric overlap factor is
\be
\label{eq:eta}
\eta_i =  \frac{\big| \int_{V_p} d^3 \v{x} ~ \v{E}_i \cdot \hat{\v{p}} \big|}{\sqrt{V  \int_V d^3 \v{x} \ (\epsilon / \bar{\epsilon}) \, E_i^2}}
~.
\ee
This definition is chosen so that $\eta_i \sim 1$ when the cavity is completely filled with dielectric polarized along $\hat{\v{p}}$ parallel to the electric field $\v{E}_i$ of the cavity mode. Below, we suppress mode subscripts to simplify notation. 

To probe the lowest possible axion masses, a cylindrical cavity can be completely filled with a dielectric with $\hat{\v{p}}$ along the cylinder's axis, which yields $\eta \simeq 0.83$ for the $\text{TM}_{010}$ mode. In Fig.~\ref{fig:coupling}, we show two concrete ways to guarantee $\order{1}$ geometric overlap for heavier axions coupled to higher resonant modes of the cavity. First, one can insert layers of another dielectric. For example, rutile caries a negligible axion-induced current, and hence does not contribute to $V_p$. Since it has a very high permittivity at cryogenic temperatures, $\epsilon \gtrsim 10^4$~\cite{parker1961static}, thin layers would suffice to preserve a large overlap factor. Alternatively, the cavity can be filled with dielectric whose spin polarization alternates in direction. In either case, the mode frequency can be coarsely tuned by changing the number of layers, and finely tuned by introducing gaps and moving the dielectric layers or endcaps along the cylinder's axis.

Such layered structures have been proposed, prototyped, operated, and tuned for haloscopes targeting the axion-photon coupling~\cite{Morris:1984nu,Sikivie:1993jm,Rybka:2014cya,electrictiger,Egge:2020hyo,Cervantes:2022yzp,Cervantes:2022epl,Lee:2022zhs}. Axions can also be effectively coupled to higher-order modes by loading cavities with dielectric wedges or cylindrical shells~\cite{Quiskamp:2020yrx,McAllister:2017ern,Kim:2019asb,QUAX:2020uxy,DiVora:2022tro,Alesini:2022lnp}. At high axion masses, scanning can become impeded by mode crowding. Many innovative approaches have been considered to avoid this issue, such as open resonators~\cite{Cervantes:2022yzp,Cervantes:2022epl}, phase-matched, coupled, or sub-divided cavities~\cite{Jeong:2017xqz,Yang:2020xsc,castcapp,Goryachev:2017wpw,Melcon:2018dba,ArguedasCuendis:2019swy,AlvarezMelcon:2020vee,CAST:2020rlf,Diaz-Morcillo:2021psa,Jeong:2017hqs,Jeong:2020cwz,Jeong:2022akg}, rod or wire metamaterials~\cite{Lawson:2019brd,Wooten:2022vpj,Bae:2022ydq}, and thin-shell geometries~\cite{Kuo:2019cps,Kuo:2020llc}. Most of these ideas can be adapted to polarization haloscopes, though some tuning mechanisms must be adjusted. For concreteness, we take $\eta = 1$, assume a cylindrical cavity with aspect ratio $L/R = 5$, and require the intermediate layers in Fig.~\ref{fig:coupling} be at least $1 \ \cm$ thick, so that there is a reasonable number to tune. This determines the mass range probed in Fig.~\ref{fig:reach}.


\section{Material Properties}
\label{sec:material}

To maximize the signal strength, we consider dielectric materials with a high density of the nuclei in \Tab{nuclei}. Unlike other approaches that require the material to be ferroelectric~\cite{Budker:2013hfa} or piezoelectric~\cite{Arvanitaki:2021wjk}, we only require the material to be insulating at low temperatures.

Some semiconducting or insulating candidate materials are nitrides $\text{XN}$~\cite{natali2013rare}, oxides $\text{XO}$, and sesquioxides $\text{X}_2\text{O}_3$ for $\text{X} = \text{Dy}, \text{Eu}, \text{Gd}$. Though many alternatives exist, these materials are simple and well-studied, and most are commercially available. For a prototype setup, we consider $\text{EuN}$ where the abundance of ${}^{153}\mathrm{Eu}$ is 52\% (see Table~\ref{tab:nuclei}). Following other proposals~\cite{Arvanitaki:2021wjk,Budker:2013hfa}, we assume complete isotope separation for a full-scale experiment, using $\text{DyN}$ where the dysprosium is entirely ${}^{161}\text{Dy}$. In both cases, the number density of rare earth atoms is $3 \times 10^{22} \ \cm^{-3}$~\cite{MaterialsProject,petousis2017high}.

The structure of the material also directly affects the strength of the signal. The most important effect, displayed in \Eq{signal_power}, is that dielectrics shield electric fields, reducing the signal power by a factor of the permittivity $\bar{\epsilon}$. For our projections we take $\bar{\epsilon} \simeq 7$, based on the static permittivity of DyN~\cite{xue2000dielectric}. This choice is conservative, as permittivity decreases at higher frequencies. 

In addition, the effective atomic EDM may be modified within a crystal, where atomic orbitals are deformed. This effect is quantified by the ``electroaxionic'' tensor defined in Ref.~\cite{Arvanitaki:2021wjk}, and calculating the tensor components requires a dedicated relativistic many-body calculation for each material. In $\text{PbTiO}_3$, two groups found suppressions of 25\%~\cite{doi:10.1063/1.4959973} and 50\%~\cite{Ludlow:2012va}, but with comparably large uncertainties. Thus, for this initial study we simply take $d_A$ to be the value for an isolated atom.

The other key material property is the dielectric loss tangent $\tan \delta$. For a cavity entirely filled with dielectric, the quality factor $Q$ of a mode obeys $1/Q = 1/Q_c + \tan \delta$, where $Q_c$ is the quality factor due to cavity wall losses. Thus, to realize a desired $Q$, one must have $\tan \delta \lesssim 1/Q$. 

At room temperature, dielectrics display high losses due to thermal phonons. However, these ``intrinsic'' losses fall steeply with temperature~\cite{gurevich1991intrinsic}, and are negligible at the cryogenic temperatures of polarization haloscopes. Instead, extrinsic losses due to defects and impurities dominate~\cite{alford2001dielectric,aupi2004microwave} and depend on crystal quality. Very low losses have been measured~\cite{braginsky1987experimental,tobar1998anisotropic,krupka1999use,krupka1999complex}, at the level of $10^{-9}$ for sapphire and $10^{-8}$ for rutile and YAG. 

These are all centrosymmetric crystals, and thereby avoid additional loss mechanisms that would appear in more complex crystals, e.g.~through acoustic phonons in piezoelectrics~\cite{gurevich1991intrinsic} or domain wall motion in ferroelectrics~\cite{liu2015losses}. The candidate materials we have listed above are also all simple centrosymmetric crystals. However, their dielectric losses are unknown, and dedicated cryogenic measurements in high-quality crystals are needed. These should be carried out at low electric field amplitudes, because high field amplitudes can mask losses due to two-level systems~\cite{martinis2005decoherence,o2008microwave,kostylev2017determination}. 

\section{Nuclear Spin Polarization}
\label{sec:spin_pol}

The current in a polarization haloscope is proportional to the fractional nuclear spin polarization $f_p$, which is $\order{1 \%}$ in thermal equilibrium in typical cavity haloscope conditions (see Table~\ref{tab:nuclei}). However, for both polarization haloscopes and other approaches~\cite{Budker:2013hfa,Arvanitaki:2021wjk} an $\order{1}$ polarization is required for optimal sensitivity. Below we describe two potential approaches to realize this.

First, one could simply subject the dielectric to a high magnetic field $B \gtrsim 10 \ \text{T}$ and ultra-low temperature. At $T = 2 \ \text{mK}$, as achieved by specialized dilution fridges~\cite{betts1989introduction,zu2022development}, ${}^{153}\text{Eu}$ nuclei possess an $\order{1}$ equilibrium polarization. For this technique, the key unknown is the time needed to thermalize the spins. At such high $B/T$, theoretical estimates suggest that it is prohibitively long~\cite{abragam1982nuclear,gonen1989nmr}, but measured spin-lattice relaxation times are much shorter than predicted~\cite{de1974dynamic,kuhns1987unexpectedly}, which could be explained by exotic relaxation mechanisms~\cite{waugh1988mechanism,phillips1988spin,vega2006spin}. Relaxation times might be further reduced by the electric quadrupole moments of the nuclei we consider, which couple more strongly to the lattice than magnetic dipole moments~\cite{abragam1961principles}, or by the addition of relaxation agents~\cite{krjukov2005brute,owers2013high}. 

Another option is frozen spin dynamic nuclear polarization (DNP), in which electrons are polarized in a few-Tesla field at $T \sim 1 \ \text{K}$, and their polarization is transferred to the nuclear spins by applying $\sim 1 \ \text{W}/\text{kg}$ of microwave power. This method achieves almost complete proton spin polarization and has been extended to heavier nuclei for NMR studies~\cite{ardenkjaer2003increase,lee2015solid,ardenkjaer2016present,budker_nmr}. It requires the sample to contain a concentration $\sim 10^{-3}$ of paramagnetic centers, produced by chemical doping or ionizing radiation. To ``freeze'' the nuclear spins, the microwave field is removed and the sample is further cooled to slow relaxation. 

This approach has been used for decades to polarize targets for particle physics experiments~\cite{Crabb:1997cy,Goertz:2002vv}; notably, the Spin Muon Collaboration at CERN produced frozen spin targets of liter scale~\cite{SpinMuon:1999uhx}. Currently, frozen spin DNP is primarily developed in nuclear physics experiments~\cite{keith2011polarized,niinikoski2020physics,targets1,targets2,targets3,targets4}. The resulting spin polarization is robust, with spin-lattice relaxation times of nearly a year observed in practice~\cite{Keith:2012ad}. For polarization haloscopes, the next step is to see how this approach can be scaled to larger volumes, while maintaining low dielectric losses.


\begin{figure}
\includegraphics[width=\columnwidth]{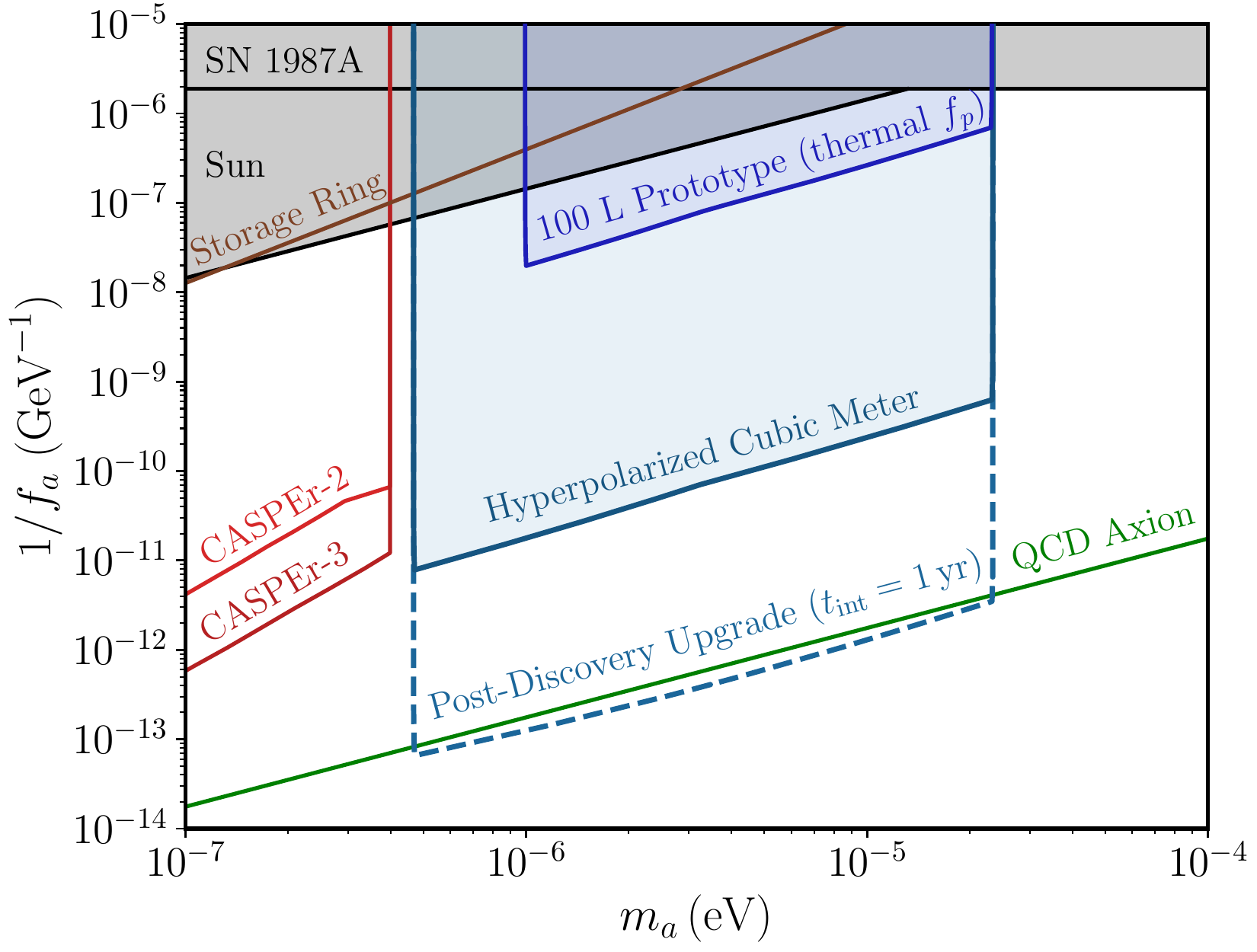}
\caption{The projected sensitivity for three benchmark polarization haloscopes (see text for details). The blue shaded regions indicate the reach of scanning setups, while the dashed blue line shows the reach for an experiment which targets a single candidate QCD axion mass. We also show the ultimate projected sensitivity of CASPEr-Electric~\cite{JacksonKimball:2017elr} and storage ring~\cite{Kim:2021pld} experiments, as well as existing constraints from the cooling of Supernova 1987A~\cite{Lucente:2022vuo} and Solar fusion processes~\cite{Hook:2017psm}. Note that these existing constraints are strictly stronger than those derived from Big Bang nucleosynthesis~\cite{Blum:2014vsa} (not shown).}
\label{fig:reach}
\end{figure}

\section{Projected Sensitivity}
\label{sec:sensitivity}

The signal-to-noise ratio is given by the Dicke radiometer equation~\cite{dicke1946measurement},
\be
 \label{eq:dicke_SNR}
\text{SNR} \simeq \frac{P_{\text{sig}}}{T_n} \, \sqrt{\frac{t_{\text{int}}}{\Delta \nu_s}}
~,
\ee
where $t_\text{int}$ is the time spent probing each axion mass, and $\Delta \nu_s = m_a / (2\pi \max(Q, Q_a))$ is the signal bandwidth. The noise temperature $T_n = T + T_{\text{amp}}$ receives comparable contributions from thermal noise, determined by the physical temperature $T$, and amplifier noise. Following Ref.~\cite{Arvanitaki:2021wjk}, we find that noise due to external vibrations or spin fluctuations is vastly subdominant at the GHz frequencies of interest, even with the inclusion of paramagnetic centers as required for DNP. Note that $Q$ is the quality factor of the cavity mode with dielectric losses included; thus, thermal noise automatically includes both the noise from electrons in the cavity walls and dielectric noise, by the fluctuation-dissipation theorem.

In \Fig{reach}, we show the projected sensitivity (corresponding to $\text{SNR} \geq 2$) for three experimental setups. The two blue shaded regions indicate scanning setups which take frequency steps of size $m_a / \min(Q, Q_a)$ with a uniform $t_{\text{int}}$, so that one $e$-fold in axion mass is scanned in one year. Following existing haloscope experiments, we assume an operating temperature of $T = 40 \ \text{mK}$~\cite{CAPP:2020utb} and an amplifier operating at the quantum limit, $T_{\text{amp}} \simeq m_a$. When thermal noise dominates, we assume the cavity is optimally overcoupled to the readout, which modestly improves the SNR by a factor of $\sqrt{T/T_{\text{amp}}}$~\cite{Berlin:2019ahk}. 

The ``prototype'' projection, shown in dark blue, is modeled on the ADMX haloscope~\cite{ADMX:2021nhd} and assumes a volume $V = 100 \ \text{L}$, quality factor $Q = 10^5$, and magnetic field $B = 8 \ \text{T}$, which produces a thermal spin polarization $f_p \simeq 5\%$ for ${}^{153}\text{Eu}$. This benchmark shows that new parameter space can be explored with minimal investment. (However, this parameter space may be in tension with the stability of white dwarfs~\cite{Balkin:2022qer}.)

The light blue projection considers a cubic meter cavity with $Q = 10^6$ and complete spin polarization, $f_p = 1$. Such an experiment would require a large dilution fridge, like those developed for other precision experiments~\cite{CUORE:2015thw,Alduino:2019xia,Hollister:2021lhg,Astone:1991ax,Cerdonio:1997hz}, and several tons of dielectric material. In other words, it would require investment comparable to ongoing WIMP dark matter searches~\cite{Boulay:2012hq,XENON:2020kmp}. Though it does not reach the canonical QCD axion line defined by Eq.~\eqref{eq:QCDmass}, it could probe orders of magnitude of unexplored parameter space, including non-minimal, mildly tuned QCD axion models which solve the strong CP problem with exponentially smaller $m_a f_a$~\cite{DiLuzio:2021pxd,DiLuzio:2021gos}.

If ADMX, CAPP, or any other GHz-frequency haloscope~\cite{Crisosto:2019fcj,DMRadio:2022pkf,lowfreq1,lowfreq2,lowfreq3} detects a signal consistent with axion dark matter, a ``post-discovery'' setup, shown in dashed blue, can probe the same mass. Since it sits at a single frequency, the SNR is enhanced by $Q_a^{1/2} \sim 10^3$ for $t_\text{int} = 1 \ \text{yr}$, as compared to a scanning experiment. We assume noise is reduced, relative to the cuber meter setup, by cooling to $10 \ \text{mK}$ and reducing amplifier noise by $3 \, \mathrm{dB}$ using demonstrated vacuum squeezing techniques~\cite{HAYSTAC:2020kwv}. We also assume a quality factor of $Q = 10^8$. To achieve this quality factor one needs a material with $\tan \delta \lesssim 10^{-8}$, which has been measured for a number of compounds. As for wall losses, one can achieve $Q_c \gg 10^8$ with a superconducting cavity, since polarization haloscopes do not require large static magnetic fields. Alternatively, the mode profile can be shaped with dielectrics, a technique which has achieved $Q \sim 10^7$ in a liter-scale copper cavity~\cite{DiVora:2022tro}. With these enhancements, a polarization haloscope has the unique ability to probe the minimal QCD axion.


\section{Discussion}
\label{sec:conclusion}

The QCD axion is an exceptional dark matter candidate, which arises automatically in theories which solve other problems of the Standard Model, with a simple and predictive production mechanism. The minimal QCD axion also has the unique advantage of possessing a defining coupling to the Standard Model, which provides a sharp target for laboratory searches. 

A polarization haloscope naturally targets higher frequencies than nuclear magnetic resonance experiments~\cite{Budker:2013hfa}. Both approaches detect the electromagnetic fields generated by spin polarized nuclei, but polarization haloscopes do not involve changes in the spin direction and hence do not require long spin coherence times. One could also target kHz to MHz frequencies with our approach by replacing the magnetic field in an LC circuit haloscope~\cite{Sikivie:2013laa,Chaudhuri:2014dla,Kahn:2016aff} with a polarized dielectric. 

We have laid out a path towards definitively probing the QCD axion with polarization haloscopes. No fundamentally new technologies are required, but many uncertainties remain. Precisely computing the signal requires expertise in theoretical nuclear, atomic, and solid state physics, while the cavity design and the selection and polarization of the material require experimental investigation. Together, such efforts may enable the next definitive search for dark matter.

\begin{acknowledgments}
We thank John Behr, Raphael Cervantes, Andrei Derevianko, Victor Flambaum, Roni Harnik, Anson Hook, Yoni Kahn, Amalia Madden, Surjeet Rajendran, Gray Rybka, Alex Sushkov, and Natalia Toro for helpful discussions. This material is based upon work supported by the U.S.~Department of Energy, Office of Science, National Quantum Information Science Research Centers, Superconducting Quantum Materials and Systems Center (SQMS) under the contract No. DE-AC02-07CH11359. Fermilab is operated by the Fermi Research Alliance, LLC under contract No.~DEAC02-07CH11359 with the United States Department of Energy. KZ is supported by the NSF GRFP under grant DGE-1656518.
\end{acknowledgments}

\bibliographystyle{utphys}
\bibliography{AxionHF}

\end{document}